\documentclass[authoryear,final,1p,times]{elsarticle}

\usepackage{amssymb}
\usepackage{amsmath}
\usepackage{booktabs}
\usepackage{longtable}
\usepackage{siunitx}
\usepackage{hyperref}
\usepackage{times}

\begin{document}

\begin{frontmatter}

\title{Contrasting Global and Patient-Specific Regression Models via a Neural Network Representation}

\author[imbi,fdm]{Max Behrens \corref{cor1}}
\cortext[cor1]{Corresponding author: max.behrens@uniklinik-freiburg.de}
\author[pneumo]{Daiana Stolz}
\author[pneumo]{Eleni Papakonstantinou}
\author[nephro]{Janis M. Nolde}
\author[math]{Gabriele Bellerino}
\author[math]{Angelika Rohde}
\author[imbi,fdm]{Moritz Hess}
\author[imbi,fdm]{Harald Binder}

\affiliation[imbi]{organization={Institute of Medical Biometry and Statistics, Faculty of Medicine and Medical Center, University of Freiburg},
            addressline={Stefan-Meier-Straße 26}, 
            city={Freiburg},
            postcode={79104}, 
            country={Germany}}
\affiliation[fdm]{organization={Freiburg Center for Data Analysis and Modeling, University of Freiburg},
            addressline={Ernst-Zermelo-Straße 1}, 
            city={Freiburg},
            postcode={79104}, 
            country={Germany}}
\affiliation[pneumo]{organization={Clinic of Pneumology, Medical Center – University of Freiburg, Faculty of Medicine, University of Freiburg},
            addressline={Killianstrasse 5}, 
            city={Freiburg},
            postcode={79106}, 
            country={Germany}}
\affiliation[nephro]{organization={Department of Nephrology, Faculty of Medicine and Medical Center, University of Freiburg},
            addressline={Hugstetter Strasse 55}, 
            city={Freiburg},
            postcode={79106}, 
            country={Germany}}
\affiliation[math]{organization={University of Freiburg, Department of Mathematical Stochastics},
            addressline={Ernst-Zermelo-Straße 1}, 
            city={Freiburg},
            postcode={79104}, 
            country={Germany}}

\begin{abstract}
When developing clinical prediction models, it can be challenging to balance between global models that are valid for all patients and personalized models tailored to individuals or potentially unknown subgroups. To aid such decisions, we propose a diagnostic tool for contrasting global regression models and patient-specific (local) regression models. The core utility of this tool is to identify where and for whom a global model may be inadequate. We focus on regression models and specifically suggest a localized regression approach that identifies regions in the predictor space where patients are not well represented by the global model. As localization becomes challenging when dealing with many predictors, we propose modeling in a dimension-reduced latent representation obtained from an autoencoder. Using such a neural network architecture for dimension reduction enables learning a latent representation simultaneously optimized for both good data reconstruction and for revealing local outcome-related associations suitable for robust localized regression. We illustrate the proposed approach with a clinical study involving patients with chronic obstructive pulmonary disease. Our findings indicate that the global model is adequate for most patients but that indeed specific subgroups benefit from personalized models. We also demonstrate how to map these subgroup models back to the original predictors, providing insight into why the global model falls short for these groups. Thus, the principal application and diagnostic yield of our tool is the identification and characterization of patients or subgroups whose outcome associations deviate from the global model.
\end{abstract}


\end{frontmatter}


\section{Introduction}

In clinical research, conclusions about potential relationships between patient characteristics and outcomes often are based on regression models. When just a single model is fitted to all patients in a dataset, this might be called a \textit{global model}, as it does not take into account that there might be some individuals for whom different values of the regression parameters would be more adequate. More specifically, there might not be just some random variability across the parameters of patients, e.g. as considered in regression modeling with random effects \citep{Pinheiro2000}, but different regions in the space spanned by the patient characteristics might require different parameters. For example, the relation of some patient characteristics to the outcome might be more pronounced for older patients with high body weight, without having a corresponding pre-defined subgroup indicator. While sticking to a global model keeps interpretation simple and is beneficial in terms of statistical stability, it would at least be useful to have some diagnostic tool for judging the potential extent of deviations from the global model. 

Some of this could be achieved by introducing interaction terms into regression models \citep{Royston2008}, but this quickly becomes unwieldy if several regression model parameters jointly depend on patient characteristics, or the number of patient characteristics that affect regression parameters is large. This challenge is particularly pertinent when dealing with data where the sample size is limited relative to the number of predictors, hindering the exploration of numerous complex interactions. To illustrate our approach to this challenge, our work utilizes a baseline cross-sectional dataset from $n=217$ patients involved in a multicenter clinical trial. For these patients, $p=76$ lung and physiological measurements were considered as predictors. This situation contrasts with big data scenarios, characterized by vast sample sizes. In such small data settings, where evaluating numerous interactions is infeasible, a more natural way for investigating regions of deviation from the global model in the space of patient characteristics is to directly consider \textit{local models}, i.e. patient-specific models that can be fitted at each point in that space and thus can adapt to individual patients or (unknown) subgroups. Technically, this can be obtained by locally weighted regression \citep{loaderOriginsLocalRegression1999,tutzLocalizedClassification2005}. 
To facilitate such patient-level diagnostics, this work introduces a novel diagnostic tool. This approach leverages a specialized local regression approach to contrast global and patient-specific models, aiming to identify individuals or subgroups for whom the global model may be insufficient.

When not aiming for a comparison to a global regression model, but directly aiming for personalized modeling, several alternative techniques have been proposed. Many of these rely on tree-based greedy approaches \citep[see][for an overview]{lipkovichTutorialBiostatisticsDatadriven2017}. Linear Model Trees (LMTs) \citep{wangInducingModelTrees1997, pottsIncrementalLearningLinear2004}, for instance, recursively split the data and fit linear models in each terminal node, yielding linear relationships on a subgroup-level. While flexible, these methods optimize splitting based on within-subgroup homogeneity without relating to a global model. This can lead to subgroups that are not relevant for understanding deviations from the global model. There are other approaches that rely on partitioning of the data that more directly target deviations from a global model, such as Exceptional Model Mining (EMM) approaches \citep{duivesteijnExceptionalModelMining2016}, or the approaches proposed by \cite{duivesteijnDifferentSlopesDifferent2012} and \cite{suttonIdentifyingDomainsApplicability2020}. However, these cannot leverage smooth variation of regression model parameters across the space spanned by patient characteristics. Other recent approaches, such as DDGroup \citep{izzoDataDrivenSubgroupIdentification2023}, identify subgroups in which a global linear model remains adequate, offering insight into where the global approach works well. However, they provide no model for the regression function outside of these stable regions, i.e. do not provide personalized models where needed. 
In contrast, local regression leverages smooth variation of regression parameters across the space of patient characteristics due to its origin from flexible non-parametric methods, in particular, weighted estimation within each kernel-defined neighborhood in the predictor space \citep{clevelandLocallyWeightedRegression1988, fanStudyVariableBandwidth1996}. Localized linear regression similarily is related to varying-coefficient models \citep{hastieVaryingCoefficientModels1993a} when considering only single covariates. 

Implementing localized regression with a larger number of patient characteristics poses challenges (e.g., $p=76$ in our application). In the original, potentially high-dimensional predictor space, data points become sparse and approximately equidistant. This makes the notion of local neighborhood and dense regions, which rely on proximity, problematic due to the curse of dimensionality  \citep{bellmanApproximationCurvesLine1961, hastieKernelMethods2001}. As \cite{loaderOriginsLocalRegression1999} demonstrated, successful localization is only feasible in a few dimensions. 

To overcome this, the core of our proposed diagnostic tool is a novel methodology to enable effective localized regression and model contrast, particularly in settings with many predictors relative to sample size. Our approach is based on the assumption that the data, despite its high dimensionality, lies on or near an underlying low-dimensional manifold. Instead of defining local neighborhoods in the original $p$-dimensional space where concepts of distance are confounded by sparsity, our method learns this manifold by mapping patient characteristics into a substantially lower-dimensional latent space ($Z \in \mathbb{R}^{n\times d}$, where $d\ll p$). It is within this learned low-dimensional latent space that we define local neighborhoods for the purpose of localized regression. Yet, we want to avoid a two-step approach, where the initial dimension reduction step is not aware of the subsequent use for local regression, and thus cannot optimize the latent representation for this task. Therefore, we propose to use an autoencoder architecture that uses artificial neural networks to learn a latent representation \citep{Hinton1993}. The loss function used for its end-to-end training is tailored to learn a latent representation that not only allows for reconstruction of the original data but is also structured to reveal meaningful local variations concerning the clinical outcome. This process aims to mitigate the curse of dimensionality by enabling localized analysis in an optimized, lower-dimensional setting where such local structures can be identified.

In Section~\ref{sec:methods}, we detail the construction of our loss function to identify a latent representation and fit localized regression models end-to-end. Section~\ref{sec:results} illustrates the proposed approach in an application to data from COPD patients, including proposals for quantities and visualizations that aid interpretation of potential deviations from global regression models. Further, we compare our proposal to the LMT approach. Section~\ref{sec:discussion} discusses implications for clinical prediction modeling, limitations, and future research directions.

\begin{figure}[!ht]
    \centering
    \includegraphics[width=\textwidth]{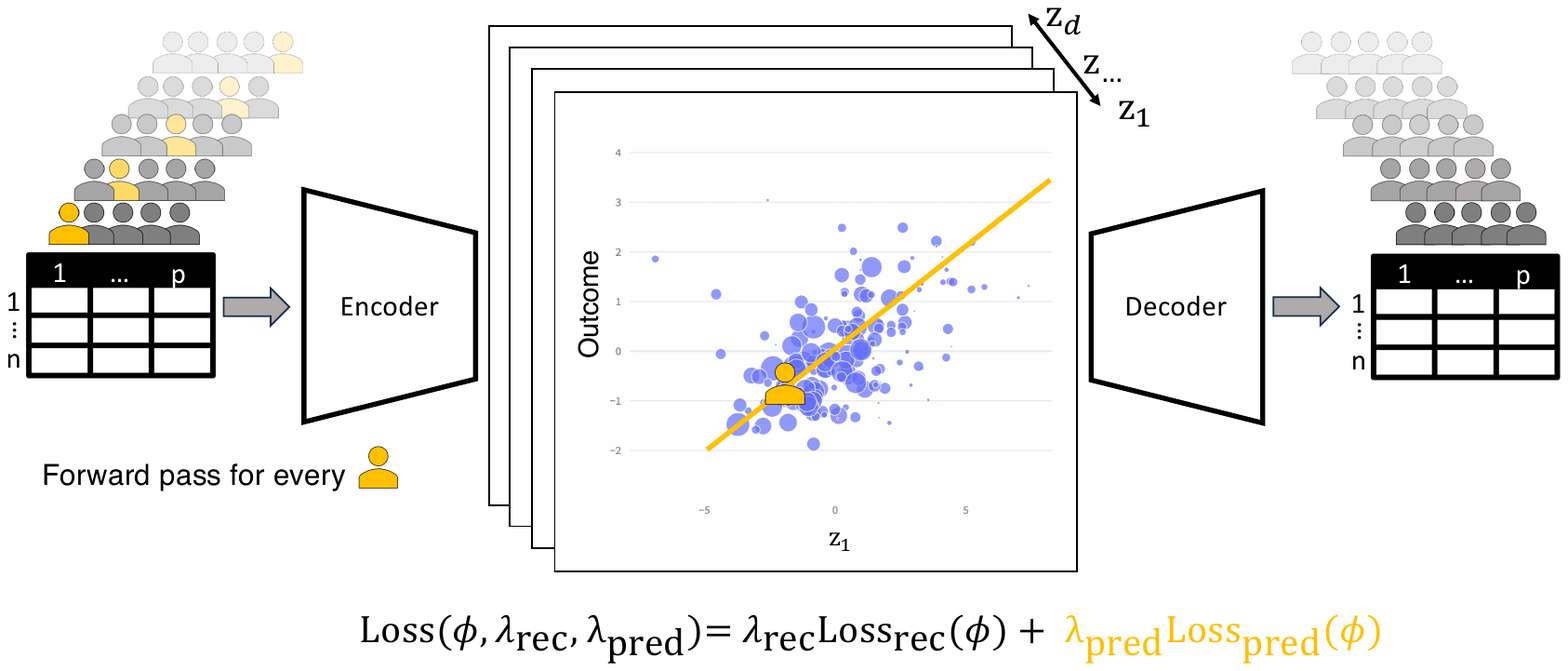}
    \caption{Schematic overview of the proposed approach. All data is passed in batches through the encoder neural network of an autoencoder to obtain values $z_1, \ldots, z_n$ of a latent representation. Localized regression is performed for each patient individually in the latent space (one exemplary individual marked in orange). The localized regression model takes the latent dimensions as predictors and the clinical outcome as target. The values of the latent representation are passed through the decoder neural network for reconstructing the original data. The loss function compromises both a reconstruction component ($\text{Loss}_{\text{rec}}$) and a component for the localized models ($\text{Loss}_{\text{pred}}$) to promote a latent representation where the outcome can be well predicted.}
    \label{fig:overview} 
\end{figure}

\section{Methods} \label{sec:methods}

\subsection{Overview of the Proposed Approach}

To assess whether a global regression model is adequate for all patients in a clinical dataset or if certain patients require localized models, we propose an approach that combines dimensionality reduction with localized regression analysis. Figure \ref{fig:overview} illustrates the core idea. The predictor matrix $\mathbf{X} \in \mathbb{R}^{n \times p}$ is mapped into a lower-dimensional latent representation $\mathbf{Z} \in \mathbb{R}^{n \times d}$ using an autoencoder, where $d \ll p$. This dimensionality reduction facilitates effective localization by creating a denser representation of the data, allowing us to capture local variations in the relationships between predictors and outcomes. Within this latent space, we fit both a global regression model and patient-specific localized regression models. By comparing the coefficients of these models, we can identify patients for whom the global model may not provide an adequate fit, indicating the need for personalized models. 

\subsection{Dimension Reduction with an Autoencoder}

We employ an autoencoder neural network to learn the deterministic mapping from the original predictor space $\mathbf{X}$ to the latent space $\mathbf{Z} = f(\mathbf{X})$. The autoencoder consists of two components: an encoder function $f: \mathbb{R}^{n\times p} \mapsto \mathbb{R}^{n\times d}$ (left trapezoid in Figure \ref{fig:overview}) and a decoder function $g: \mathbb{R}^{n\times d} \mapsto \mathbb{R}^{n\times p}$ (right trapezoid in Figure \ref{fig:overview}) leading to the reconstruction $g(f(\mathbf{X}))=\mathbf{\hat{X}}$, both implemented as multi-layer perceptrons (MLPs). These MLPs consist of three fully connected layers each. The encoder reduces the dimensionality by capturing the most relevant features of the data, while the decoder attempts to reconstruct the original input from the latent representation \citep{rumelhartLearningRepresentationsBackpropagating1986}. Each layer $l$ of the encoder and decoder computes an output vector $h^{(l)}$ based on the output of the preceding layer $h^{(l-1)}$ using the transformation:

$$
\mathbf{h}^{(l)} = a^{(l)} \left( \mathbf{W}^{(l)} \mathbf{h}^{(l-1)} + \mathbf{b}^{(l)} \right).
$$

In this formulation, $h^{(0)}$ for the first layer of the encoder is the original input data $\mathbf{X}$. $\mathbf{W}^{(l)}$ and $\mathbf{b}^{(l)}$ are the weight matrix and bias vector of layer $l$, respectively. This is followed by elementwise application of a non-linear activation function $a^{(l)}$, specifically we use a hyperbolic tangent (tanh). The result is a vector that serves as input for the next layer. The output of the final layer of the encoder constitutes the latent representation $\mathbf{Z}$. The decoder then takes $\mathbf{Z}$ as its input, and the output of its final layer is the reconstructed data $\hat{\mathbf{X}}$.

\subsection{Localized Regression}

To capture potential patient-specific divergence from the global model, we fit localized regression models for each patient within the latent space depicted in the center of Figure \ref{fig:overview}. Localization is achieved by weighting neighboring data points based on their proximity to the target point $\mathbf{z_i} \in \mathbb{R}^d$, where $\mathbf{z_i}$ represents the latent row vector for the $i$-th observation. We base our implementation on the work from \cite{tutzLocalizedClassification2005}. 

Each localized regression model is fitted to an observation $i$ (with latent vector $\mathbf{z_i}$) by weighting nearby observations $j$ (with latent vectors $\mathbf{z_j}$) based on their proximity in the latent representation $\mathbf{Z}$. The weights, $w_{ij}^{(k)}$, depend on the Euclidean distance between the target observation's latent vector $\mathbf{z_i}$ and that of a neighboring observation $\mathbf{z_j}$. This relationship is governed by a kernel function $K(\cdot)$, defined as:

$$
w_{ij}^{(k)} = K\left(\frac{\|\mathbf{z_i} - \mathbf{z_j}\|_2}{d_k(\mathbf{z_i})}\right).
$$

In this formula, $\|\mathbf{z_i} - \mathbf{z_j}\|_2$ represents the Euclidean norm between the latent vectors $\mathbf{z_i}$ and $\mathbf{z_j}$ and $d_k(\mathbf{z_i})$ is a locally adaptive kernel width parameter. 

The kernel width $d_k(\mathbf{z_i})$ is determined by the distance to the $k$-th nearest neighbor $z_{(k)}$ of $\mathbf{z_i}$. This means $d_k(\mathbf{z_i}) = \|\mathbf{z_i} - \mathbf{z_{(k)}}\|_2$. The order $k$ of the nearest neighborhood is a flexible parameter in the algorithm, allowing the method to adapt to different levels of local data density. In denser regions, $k$ can be larger, leading to smoother models, while in sparser regions, a smaller $k$ might be more appropriate to maintain localized model flexibility. Various kernel functions $K(\cdot)$ can be used to define the weights. In our investigations, we primarily use a Gaussian kernel, defined as:

$$
K_G(x) = \exp\left(-\frac{x^2}{2\sigma^2}\right),
$$

where  $\sigma$ is a bandwidth parameter that controls the spread of the kernel. The Gaussian kernel is preferred for its smoothness and its property of assigning non-zero weights to all observations. This latter characteristic can be beneficial for the stability of local models, particularly in sparser regions of the latent space.
Furthermore, when this kernel function $K_G(\cdot)$ is used to determine the specific weight $w_{ij}^{(k)}$ assigned to an observation $\mathbf{z_j}$ in the context of fitting a localized model for a target observation $\mathbf{z}_i$, its mathematical form (an exponential function) ensures that $w_{ij}^{(k)}$ is always non-zero for any finite distance $\|\mathbf{z}_i-\mathbf{z}_j\|_2$. These calculated weights decrease exponentially as the (scaled) distance between $\mathbf{z}_i$ and $\mathbf{z}_j$ increases. Preliminary investigations with the Tricube kernel indicated that its property of assigning zero weights to points outside a specific bandwidth led to less stable local models in our setting.

\subsection{Loss Function for End-to-End Training} \label{sec:comp_loss}

Our approach integrates a global model and localized regression models within a unified, low-dimensional latent space constructed by an autoencoder network. To achieve this, we use a composite loss function that enables the end-to-end training of the autoencoder and the localized regression model. This loss function reflects two key goals: (1) reconstructing the input data, and (2) focusing on predictors associated with the outcome to ensure the latent representation prioritizes clinically relevant predictors for meaningful local modeling.

The first component is the reconstruction loss, $\text{Loss}_{\text{rec}}(\boldsymbol{\phi})$, which ensures that the autoencoder can reconstruct the original input data from the latent representation. Specifically, we measure how well the decoder output matrix $\mathbf{\hat{X}}$ matches the original input matrix $\mathbf{X}$ using the mean squared error (MSE):

$$
\text{Loss}_{\text{rec}}(\boldsymbol{\phi}) = \frac{1}{n \cdot p} \| \mathbf{X} - \mathbf{\hat{X}} \|_F^2.
$$

Here, $\| \mathbf{X} - \mathbf{\hat{X}} \|_F^2$ denotes the Frobenius norm of the difference between $\mathbf{X}$ and $\mathbf{\hat{X}}$. The functions $f(\cdot; \boldsymbol{\phi})$ and $g(\cdot; \boldsymbol{\phi})$ are the encoder and the decoder, respectively, and $\boldsymbol{\phi}$ represents the parameters (weights and biases) of the autoencoder. Minimizing $\text{Loss}_{\text{rec}}(\boldsymbol{\phi})$ encourages the latent representation $\mathbf{Z}$ to retain the meaningful information from the original data which are necessary for reconstruction.

The second component ensures that the latent representation locally preserves the relationship between the predictors and the clinical outcome at the patient level. This encourages the latent space to focus on predictors that can predict the outcome. For each patient $i$, we fit a localized regression model and compare it to a corresponding localized null model (intercept-only) that uses the same patient-specific weighting. We then average this comparison across all patients:

$$
\text{Loss}_{\text{pred}}(\boldsymbol{\phi}) = \frac{1}{n}\sum_{{i=1}}^{n} -\log\left(\frac{\mathcal{L}(\boldsymbol{\hat{\beta}}_i \mid \mathbf{Z})}{\mathcal{L}(\boldsymbol{\hat{\beta}}_{0,i} \mid \mathbf{Z})}\right).
$$

Here, $\mathcal{L}(\beta_i \mid \mathbf{Z})$ is the Gaussian likelihood derived from a patient-specific localized weighted regression model. This model is fitted for each patient $i$ using the $d$ dimensions of the latent representation $\mathbf{Z}$ as predictors for the clinical outcome. Specifically, for each patient $i$, the local coefficient vector $\boldsymbol{\beta}_i = (\beta_{i0},\beta_{i1},\dots,\beta_{id})^T$ is estimated by minimizing the patient-specific weighted sum of squared errors: 

$$
\hat{\boldsymbol{\beta}}_i = \arg\min_{\boldsymbol{\beta}} \sum_{j=1}^{n} w_{ij}^{(k)} \left( y_j - \left( \beta_0 + \sum_{l=1}^{d} \beta_l z_{jl} \right) \right)^2
$$

where $\mathbf{y}_j$ is the clinical outcome for patient $j$ and $\mathbf{z}_j=(z_{j1},\dots,z_{jd})$ is the latent vector for patient $j$. The Gaussian likelihood $\mathcal{L}(\boldsymbol{\hat{\beta}}_i \mid \mathbf{Z})$ is then formulated based on this localized weighted least squares fit, assuming normally distributed errors.

Similarly, $\mathcal{L}(\boldsymbol{\hat{\beta}}_{0,i} \mid \mathbf{Z})$ is the Gaussian likelihood of the corresponding localized weighed null (intercept-only) linear model for patient $i$. The coefficient $\hat\beta_{0,i}$ is estimated by this:

$$
\hat\beta_{0,i}
= \arg\min_{\beta_0}
\sum_{j=1}^n w_{ij}^{(k)}\bigl(y_j - \beta_0\bigr)^2.
$$

This fitting process incorporates the same patient-specific weights $w_{ij}^{(k)}$ and assumes normally distributed errors. Minimizing $\text{Loss}_{\text{pred}}(\boldsymbol{\phi})$ thus encourages the latent representation to preserve meaningful outcome-related patterns at the patient level.

We combine these two components into a single objective function as depicted in Figure \ref{fig:overview}:

$$
\text{Loss}(\boldsymbol{\phi}, \lambda_{\text{rec}}, \lambda_{\text{pred}}) = \lambda_{\text{rec}} \,\text{Loss}_{\text{rec}}(\boldsymbol{\phi}) + \lambda_{\text{pred}} \,\text{Loss}_{\text{pred}}(\boldsymbol{\phi}),
$$

where $\lambda_{\text{rec}}$ and $\lambda_{\text{pred}}$ control the trade-off between reconstructing the original data and learning a latent representation that preserves local relationships to the outcome.
By jointly optimizing these objectives, we obtain a latent space tailored for patient-level diagnostics, enabling us to determine when the global model is sufficient and when local models offer improvements. It is important to note that due to the inclusion of non-linear transformations, the overall loss function is generally non-convex.

\subsection{End-to-end Training Process}

We train the autoencoder and the localized regression model parameters simultaneously as illustrated in Figure \ref{fig:overview} which is referred to as end-to-end training. This is achieved using a single, composite loss function (detailed in Section \ref{sec:comp_loss}), allowing error gradients from the final prediction task to inform and refine the parameters of the initial dimension reduction stage, ensuring the learned latent representation is optimized for all specified objectives.

This is leveraged by automatic differentiation software, specifically PyTorch, to compute gradients of the loss function with respect to the autoencoder parameters $\boldsymbol{\phi}$. Since both $\mathcal{L}(\beta_i \mid \mathbf{Z})$ and $\mathcal{L}(\beta_{\text{null}, i} \mid \mathbf{Z})$ depend on $\mathbf{Z}$, and $\mathbf{Z}$ depends on $\boldsymbol{\phi}$, the entire training process is end-to-end. To optimize the loss function, we use the Adam optimizer with an initial learning rate of $\alpha = 0.0001$, which is expected to converge to a local optimum. The training process involves iterating over the dataset multiple times. An epoch is defined as one complete pass through the entire training dataset. Convergence was practically assessed by monitoring the reconstruction loss over epochs and training was stopped when the reconstruction loss no longer showed meaningful improvement, which occured at 300 epochs in our application.

Given the non-convex nature of the loss function, different random initializations of the autoencoder weights can lead to different local optima. To evaluate the stability and robustness of our findings to such variations, we conducted a seed-perturbation study. This involved running the entire analysis pipeline 15 times, each with a different random seed for weight initialization. The variability of key performance metrics (reported in Section 3.2) and the stability of patient rankings based on their deviation from the global model (particularly for the most diverging individuals) were then assessed.

All analyses were performed using Python (version 3.13) with libraries PyTorch (version 2.6) \citep{paszkePyTorchImperativeStyle2019} for neural network implementation, scikit-learn (version 1.6) \citep{scikit-learn} for regression and clustering, and NumPy (version 2.2) \citep{harris2020array} for numerical computations. The code used in this study is available on GitHub (github.com/-maxjonasbehrens/AEnabledLoReg). Details on the autoencoder architecture and chosen hyperparameters can be found in the Appendix.

\subsection{Balancing Global and Local Models in the Learned Latent Representation}

After training an autoencoder to learn a low-dimensional latent representation and fitting localized regression models, we use this latent space to contrast the local regression models with a global regression model. We label each latent dimension based on its strongest associations with the original predictors (see Appendix for details on the latent correlation matrix and naming process).

To identify patients that are not described well by the global regression model, we compare their fitted local regression coefficients to those of a single, global linear regression model. This global model is fitted once using all training data, with the learned latent dimensions as predictors and the clinical outcome as the target. Consequently, when a patient's local coefficient for a specific latent dimension falls outside this global model's confidence interval, we employ this as a heuristic rule to flag the patient as potentially deviating. For this, we refer to the standard confidence intervals derived from this global fit only. This approach serves as an exploratory tool to identify individuals for whom the global model might be insufficient. Patients who are flagged as deviating are then further characterized. Those who exhibit a common pattern of deviation---specifically, their local coefficients differing from the global model's coefficient for the same latent dimension and in the same direction---are collected together. Each such collection of patients is termed a subgroup in our analysis. This approach does not assume a priori that multiple distinct subgroups exist but allows for the identification of one or more such subgroups if different consistent patterns of deviation are present in the data.

After identifying patient subgroups in the learned latent space, we characterize these subgroups using the original patient predictor variables ($\mathbf{X}$). This post-hoc analysis, which helps in understanding their clinical profiles, involves two main approaches: examining interactions effects and comparing profiles based on the average Z-scores of the original, standardized predictors.

To generate interpretable Z-score profiles (e.g. Figures \ref{fig:group1}C and \ref{fig:group2}C), we calculate the average Z-score for each of the $p$ original predictors for patients within a specific subgroup. Since these predictors were standardized before training the autoencoder, comparing a subgroup's average Z-scores to the overall population of zero reveals which initial patient characteristics are particularly prominent or diminished. This process utilizes the original predictor values, distinct from the latent space variables.

For a more digestible presentation, especially with numerous original predictors (e.g. $p=76$ in our application), the predictors are grouped into fewer clusters (e.g., 21 in our case) using hierarchical clustering based on their correlations. These predictor clusters, labeled descriptively (see Appendix for clustering and labeling details), are solely for summarizing subgroup characteristics more effectively. Importantly, this clustering of original predictors is purely for interpretative aid and does not influence the patient subgroup identification, which occurs in the latent space.

\subsection{Comparison with Benchmark Dimensionality Reduction Models}

To evaluate the specific contribution of our outcome-guided latent representation, we compared its utility for outcome prediction against two standard dimensionality reduction techniques, a principal component analysis (PCA) and a autoencoder without the prediction loss term ($\text{Loss}_{\text{pred}}$).

For PCA, the original 76 standardized predictors from the training set were projected onto the first four principal components. This dimensionality was chosen to match the latent dimension of our proposed method, ensuring a fair comparison.

For the non-linear benchmark, we implemented a regular autoencoder. This autoencoder utilized the same architecture as our proposed model (three fully connected layers in both the encoder and decoder, with hyperbolic tangent activation functions, as detailed in Section 2.2 and Appendix A.1) and was also trained to reduce the 76 predictors to four latent dimensions. However, critically, this regular autoencoder was trained optimizing only the reconstruction loss ($\text{Loss}_{\text{rec}}$), without the $\text{Loss}_{\text{pred}}$ component. This allows for a direct assessment of the benefit gained by incorporating outcome-related associations into the latent space learning process.

The primary metric for comparing the informativeness of the latent spaces generated by our proposed method, PCA, and the regular autoencoder was the explanatory power for the clinical outcome. For each of the three methods, a global regression model was fitted using their respective four latent dimensions (or principal components) as predictors for outcome on the training data. The $R^2$ values from these models was then compared.

\subsection{Standard Stepwise Regression with Interaction Search}

To further contextualize the contributions of our approach as a diagnostic tool, we compared its findings with those from a traditional statistical modeling strategy aimed at building a global regression model with main effects and pairwise interactions from the original predictor space. This benchmark approach involved a multi-step procedure performed on the training data ($n=173$).

Each of the 76 original predictors  was first assessed univariately for its association with the SGRQ outcome using a linear regression model. Different thresholds of p-values were tested for predictors to be carried forward. Starting with the variables selected in the univariate step, a backward elimination procedure was applied to select a subset of main effects. In each step, the variable whose removal resulted in the largest improvement (decrease) in the AIC was eliminated, provided this improvement exceeded a threshold of 1. The process stopped when no further variable removal met this criterion. Using the main effects identified from backward elimination, a forward selection procedure was employed to identify significant pairwise interactions. Potential interaction terms were added one at a time to the main effects model. The interaction that resulted in the largest AIC improvement (decrease) was added to the model, provided this improvement exceeded a threshold of 2. This continued until no interaction term could be added that met the AIC improvement criterion.

\section{Results} \label{sec:results}

\subsection{Clinical Data from COPD Patients}

We evaluated our method on clinical data from the PREVENT study \citep[][]{stolzIntensifiedTherapyInhaled2018}, which investigated whether intensified inhaled therapy with LABA/ICS could reduce exacerbations in patients with moderate to severe chronic obstructive pulmonary disease (COPD). The trial enrolled 450 patients, but we focused on 217 patients who did not report an upper respiratory tract infection (URTI) during the study. This subset allowed us to examine baseline relationships without the added complexity of acute infection. Our outcome was the St. George’s Respiratory Questionnaire (SGRQ) score. The SGRQ is a standard instrument measuring health-related quality of life in patients with respiratory conditions, with scores ranging from 0 to 100. Higher SGRQ scores indicate greater impairment. In the PREVENT study, the SGRQ score used as the outcome in our analysis was measured at the six-month follow-up visit. The SGRQ score, prior to standardization, had a mean of 37.7 and a standard deviation of 19.1. We used 76 lung and physiological measurements as predictors and split the data into training (173 patients) and test (44 patients) sets. Before training, we standardized both the predictors and the outcome to a mean of 0 and a standard deviation of 1, ensuring that scale differences did not skew the learning process and subsequent analysis. A detailed explanation of the pre-processing steps of the data can be found in the Appendix.

\subsection{Establishing and Evaluating the Global Latent Model}

The specific findings and visualizations presented in this section are derived from a single run deemed most representative from the 15 seed-perturbation experiments (see Section 2.5), which was chosen based on the median reconstruction loss across all seeds.

Within our proposed autoencoder-based framework, we aim to determine whether a single global model---which is fitted on the whole training data using the learned latent dimensions as predictors---can adequately represent all patients or if subsets exhibited poor fit that could be improved by the localized regression models operating within this same latent space. To directly compare a global model and patient-specific localized models, we used our method to reduce the original predictor space to a four-dimensional latent representation, allowing both model types to be fit in the same low-dimensional space. Specifically, we use an autoencoder with three fully connected layers in both the encoder and decoder, the first two followed by a hyperbolic tangent activation. With 76 predictors, the encoder’s input layer has 76 nodes. The second layer has 64 nodes, and the third layer has 16 nodes that output four latent dimensions. The decoder mirrors this setup to reconstruct the data from the four-dimensional latent space.
The selection of four latent dimensions was determined experimentally by evaluating a range of dimensionalities (e.g., 2 to 10). This dimensionality provided a balance between maintaining acceptable data reconstruction accuracy, ensuring the stability of the subsequent localized regression models by avoiding excessive sparsity, and yielding interpretable latent dimensions able to identifying meaningful patient subgroups. We then fitted the global model using the four latent dimensions as predictors and the SGRQ score as the outcome variable.

Before detailing the results from the representative seed, we report on the stability of the modeling framework across the 15 different random initializations. The average training reconstruction loss was 0.41 (SD 0.00), and the average test reconstruction loss was 0.46 (SD 0.00). The global ordinary least squares (OLS) model, when fitted to the latent dimensions derived in each run, yielded an average $R^2 = 0.40$ (SD 0.03). These results, particularly the standard deviations, indicate consistent performance in data reconstruction and global model fit across different initializations. The specific results detailed below are from the representative seed, which achieved a training reconstruction loss of 0.41, a test reconstruction loss of 0.46, and whose global latent model (Table 1) explained $R^2=0.41$ of the variance in the SGRQ outcome.

The results shown in Table \ref{tab:global_model} show positive coefficients for all four latent dimensions predicting the SGRQ score for the representative seed. The 'Airflow Obstruction \& Diffusion Capacity' dimension's positive coefficient ($\beta=0.38$) intuitively links greater obstruction with poorer health. The other three dimensions, 'Dynamic Vital Capacity' ($\beta=0.38$), 'Static Lung Volumes' ($\beta=0.36$), and 'Gas Trapping \& Hyperinflation' ($\beta=0.11$), are all inversely related to their underlying clinical measurements (see Appendix). Therefore, their positive coefficients correctly indicate that factors like reduced lung capacity, lower volumes, and increased gas trapping correspond to worse health. Of these, only the coefficient for 'Gas Trapping \& Hyperinflation' was not statistically significant.

\begin{table}[t!]
\centering
\caption{Regression Results for the Global Latent Model Predicting SGRQ Score}
\label{tab:global_model}
\begin{tabular}{lS[table-format=-1.4]S[table-format=-1.4]S[table-format=-1.4]S[table-format=-1.4]}
\toprule
 & \textbf{Coefficient} & \multicolumn{2}{c}{\textbf{95\% Confidence Interval}} \\
\cmidrule(lr){3-4}
 &  & \multicolumn{1}{c}{Lower} & \multicolumn{1}{c}{Upper} \\
\midrule
Intercept                  & 0.04 \\
Airflow Obstruction \& Diffusion Capacity   & 0.38  & 0.28  & 0.49 \\
Dynamic Vital Capacity      & 0.38 & 0.27 & 0.49 \\
Static Lung Volumes  & 0.36 & 0.26 & 0.47 \\
Gas Trapping \& Hyperinflation       & 0.11 & -0.01 & 0.22 \\
\midrule
$R^2$ & \multicolumn{3}{c}{0.41} \\
\bottomrule
\end{tabular}
\end{table}

\subsection{Identifying Patients Who May Need Personalized Models}

Having established the global model, we next examined how well it represented each individual patient. To do this, we fit localized regression models to each patient and then compared these patient-specific coefficients to those from the global model. We visualized these comparisons in the learned latent space that allowed us to see where patients diverged.

\begin{figure}[!ht]
    \centering
    \includegraphics[width=\textwidth]{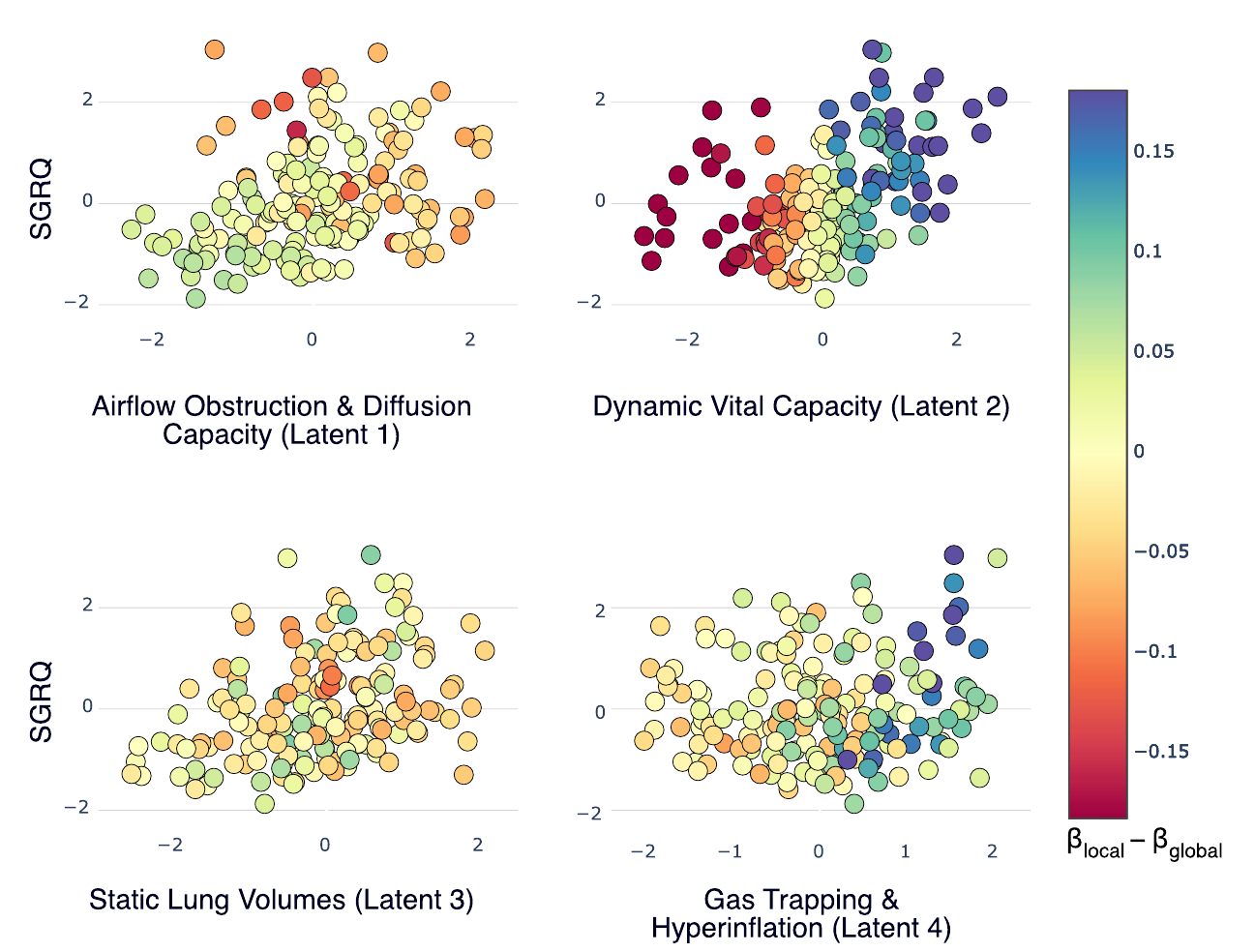}
    \caption{Visualization of patient-specific deviations from the global regression model in the learned latent space. Each subplot displays patients according to their SGRQ score and their value on one of the four latent dimensions: (1) Airflow Obstruction \& Diffusion Capacity, (2) Dynamic Vital Capacity, (3) Static Lung Volumes, and (4) Gas Trapping \& Hyperinflation. The color of each point represents the difference between the patient's local coefficient ($\boldsymbol{\beta}_{\text{local}}$) and the global model's coefficient ($\boldsymbol{\beta}_{\text{global}}$) for that dimension. Yellow indicates agreement with the global model, while blue and red hues signify positive and negative deviations, respectively. Distinct patterns of deviation, which motivate the subsequent subgroup analysis, are most apparent for Dynamic Vital Capacity (Latent 2) and Gas Trapping \& Hyperinflation (Latent 4).}
    \label{fig:latent} 
\end{figure}

To identify patients for whom the global model might be inadequate, we visualized the difference between each patient's local regression coefficient and the global coefficient for each latent dimension (Figure \ref{fig:latent}). The visualization confirms that the global model was adequate for the majority of patients, who show minimal deviation. However, we observed two distinct patterns of strong, consistent deviation from the global model, suggesting that meaningful subgroups exist. These patterns were most prominent in the second and fourth latent dimensions, providing a clear rationale for a more focused subgroup analysis.

\begin{figure}[!ht]
    \centering
    \includegraphics[width=\textwidth]{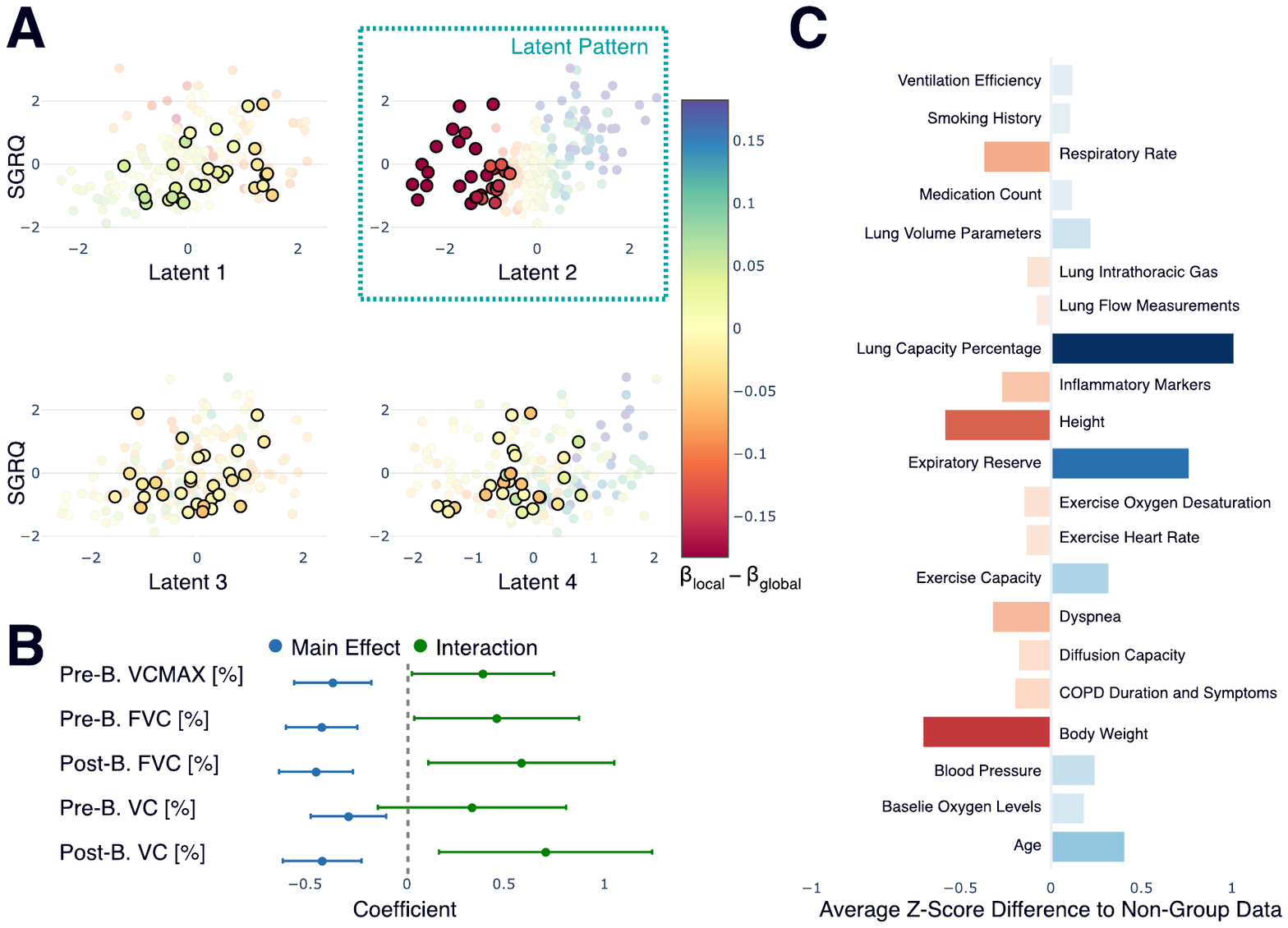}
    \caption{Characterization of Subgroup 1 ($n=30$), defined by a distinct pattern of deviation from the global model. (A) The latent space plots show this group's deviation is concentrated in the second latent dimension. (B) An interaction analysis of the original predictors most associated with this latent dimension reveals significant interaction effects, confirming a differential relationship for this subgroup. B. refers to bronchodilator in the Figure. (C) The Z-score profile shows that, compared to the rest of the cohort, these patients are characterized by lower body weight, higher lung capacity percentage, and higher expiratory reserve.}
    \label{fig:group1} 
\end{figure}

The first subgroup, comprising 30 patients, was identified by a consistent deviation pattern primarily along the second latent dimension (Figure \ref{fig:group1}A). To understand the clinical meaning of this deviation, we analyzed the original predictor variables. An interaction analysis confirmed that the relationship between key lung function measures (such as VCMAX and FVC) that are most associated with the second latent dimension and the outcome differs significantly for this subgroup compared to the rest of the cohort (Figure \ref{fig:group1}B). A profile of their baseline characteristics revealed that these patients tend to have a lower body weight, have higher lung capacity percentages, and have a higher expiratory reserve (Figure \ref{fig:group1}C). These distinct clinical features can help explain why the global model's assumptions do not fully hold for this specific group of patients.

To quantify the benefit of local modeling for the identified subgroups, we compared the Root Mean Squared Error (RMSE) for patients inside ($n=30$) and outside ($n=143$) the first subgroup. For patients within this subgroup, the local models reduced the RMSE from 0.83 (global model) to 0.73, an improvement of 0.10. In contrast, for the remaining patients, the local models provided a smaller benefit of 0.06 (an RMSE reduction from 0.73 to 0.68). A nearly doubled benefit for the subgroup confirms that our method successfully identifies patients for whom a personalized model is most advantageous.

\begin{figure}[!ht]
    \centering
    \includegraphics[width=\textwidth]{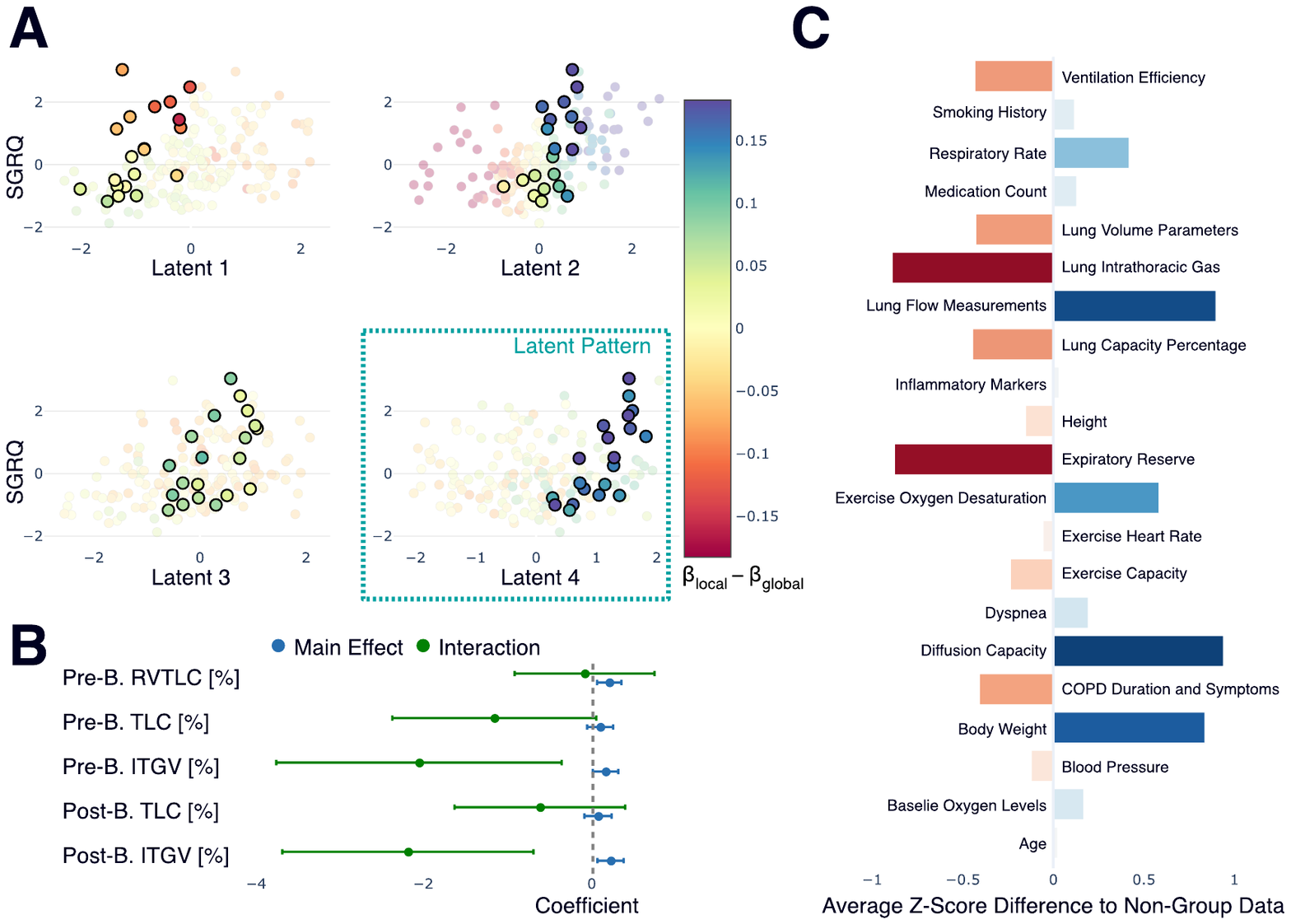}
    \caption{Characterization of Subgroup 2 (n=20), defined by a deviation pattern concentrated in the fourth latent dimension. (A) The latent space plots, with the fourth dimension highlighted, show the specific pattern defining this group. (B) An interaction analysis of associated original predictors reveals significant interaction effects for measures of intrathoracic gas volume (ITGV), indicating a differential relationship for these patients. (C) The Z-score profile shows that this subgroup is characterized by higher lung flow measurements, lower expiratory reserve, and higher diffusion capacity.}
    \label{fig:group2} 
\end{figure}

A second subgroup of 20 patients was identified, defined by a deviation pattern concentrated in the fourth latent dimension corresponding to gas trapping and hyperinflation. An analysis of the original predictors confirmed that the relationships for intrathoracic gas volume (ITGV) are significantly different for this group (Figure \ref{fig:group2}B). Clinically, these patients are distinguished by higher lung flow measurements, lower expiratory reserve, and higher diffusion capacity compared to the rest of the cohort (Figure 4C). This distinct physiological profile provides a basis for why the global model does not fully capture the predictor-outcome relationship for these individuals.

The extent to which the global model fails this subgroup and the corresponding benefit of a local approach is confirmed by a quantitative analysis of predictive error. For these 20 patients, local modeling reduced the RMSE from 0.99 to 0.83, an improvement of 0.16. This benefit is more than three times greater than the 0.05 improvement seen for patients outside this group, confirming that a personalized model is indeed more advantageous for these individuals.

Finally, to ensure these subgroups were not artifacts of a single model initialization, we assessed their stability across the 15 seeds. We evaluated the consistency with which patients were identified as top deviators by measuring the standard deviation of their deviation rank across the runs. Both subgroups showed strong stability. The second subgroup, associated with gas trapping, was particularly stable, with a mean standard deviation in rank of 4.56. The first subgroup was also consistently identified, showing a mean standard deviation in rank of 10.01. This stability analysis confirms that the identified subgroups represent consistent, reproducible patterns in the data rather than random chance.

\subsection{Comparison of Results to Test Data}

After establishing our findings with the training set (173 patients), we examined how well they held up in a separate test set (44 patients). Although the test set was too small to form distinct groups on its own, we could still map these patients onto the latent space defined by the training data. Within the latent space, the test patients generally followed the same patterns as the training patients, showing no major shifts in coefficient differences.


\begin{figure}[!ht]
    \centering
    \includegraphics[width=\textwidth]{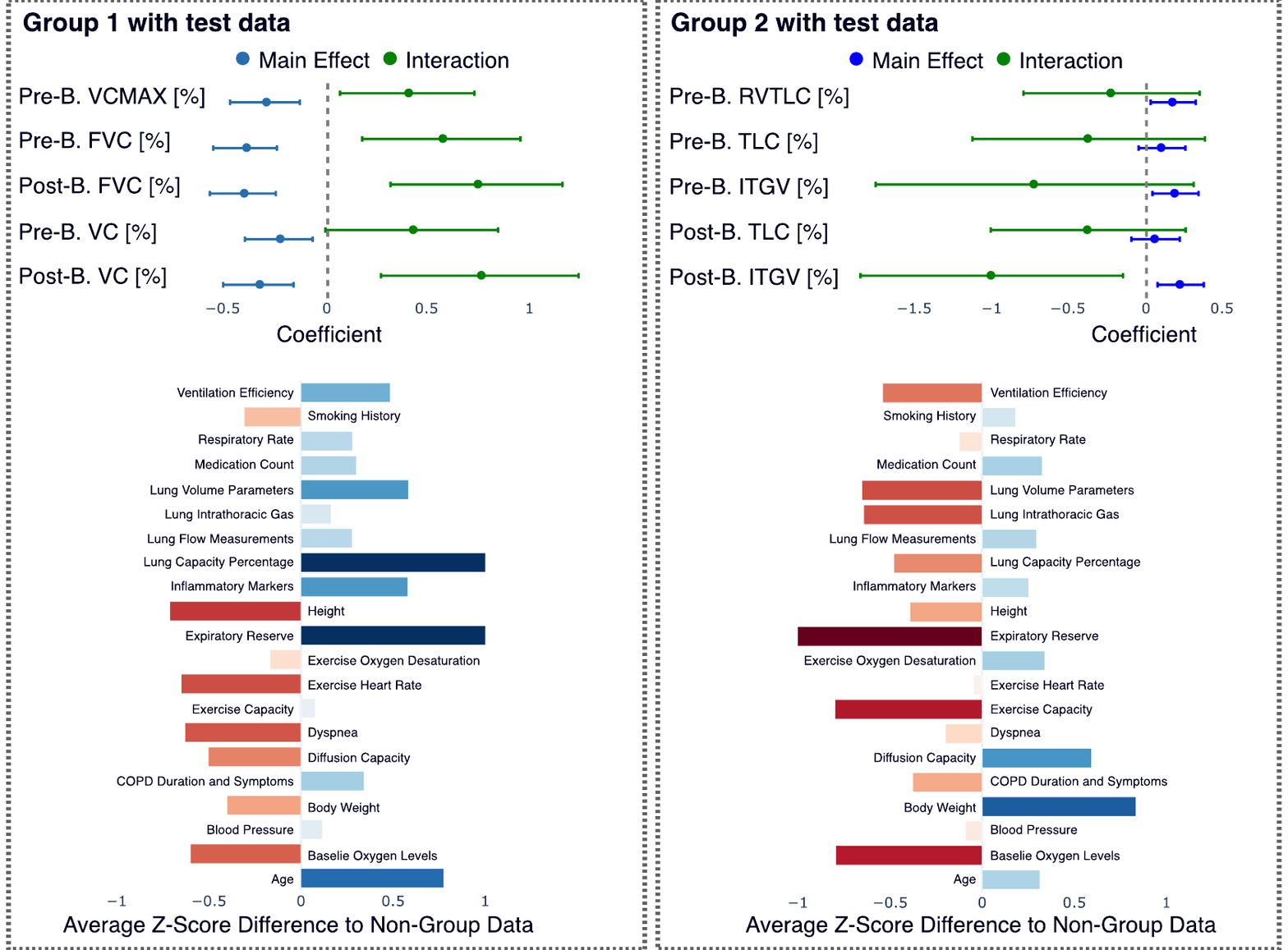}
    \caption{Test patients assigned to Group 1 (n=6) and Group 2 (n=6) show similar patterns to training patients. Regression coefficients and interaction terms remained consistent, with minor changes due to increased sample size. Z-score profiles confirm that key characteristics are largely preserved in test patients. B. refers to bronchodilator in the Figure.}
    \label{fig:groups_test} 
\end{figure}

The key characteristics of these subgroups were preserved in the test set, as shown in Figure \ref{fig:groups_test}. When the test patients were added, the interaction effects for the primary associated predictors remained significant. Furthermore, the clinical Z-score profiles of the test patients aligned with their training set counterparts. Test patients assigned to Group 1 also showed lower body weight and higher expiratory reserve, while those in Group 2 exhibited higher lung flow measurements and diffusion capacity. Despite the small sample size and expected variability, this consistency demonstrates that the learned latent space and the corresponding subgroup definitions generalize to new, unseen data.

\subsection{Comparison to a Stepwise Approach for Detecting Interaction Effects}

To contrast our local modeling approach with a traditional statistical method, we performed a standard stepwise regression procedure on the original 76 predictors. We varied the initial univariate screening p-value threshold across 0.05, 0.10, and 0.157 to generate a set of global models for the SGRQ score. The AIC improvement thresholds for main effect elimination (backward) and interaction addition (forward) remained fixed at 0 and 2, respectively. This allowed us to investigate how the selection of main effects and interactions by the standard approach changes with initial stringency and how these compare to the insights from our localized regression method.

The stepwise procedure, particularly with a lenient initial screening threshold ($p = 0.10$), produced a global model with high predictive power ($R^2 = 0.59$) and identified several interactions involving lung volume and capacity variables, such as COPD symptoms with VCMAX and FVC with FEV1. This shows a clear overlap with the variables that define our first subgroup, suggesting that both methods have a similar sensitivity for detecting these specific effect modifications when the initial variable selection is broad enough.

However, a key limitation of the stepwise method is its reliance on this initial screening. Predictors must demonstrate a significant global main effect to be considered for the model. This creates a risk of prematurely filtering out variables that are crucial for defining a subgroup but whose effects are diluted across the entire population.

Our local modeling approach overcomes this drawback, as it does not require global significance to detect localized effects. This is best illustrated by the discovery of our second subgroup, which is defined by interactions related to gas trapping and intrathoracic gas volume (ITGV). These localized relationships were uncovered by our method even though the involved predictors and their specific interactions were not selected by the stepwise procedure. This demonstrates our method's ability to serve as a diagnostic tool, revealing clinically relevant subgroups with distinct effect modifications that might otherwise be missed by traditional global model-building techniques. The full models identified by the stepwise approach can be found in the Appendix.

\subsection{Comparison with PCA and Regular Autoencoder}

To demonstrate the advantages of our proposed autoencoder, we compared its ability to generate an outcome-informative latent space against two standard methods: principal component analysis (PCA) and a regular autoencoder trained solely for reconstruction. For each method, we used the resulting four-dimensional space to fit a global regression model for the SGRQ score, reporting the average performance over 15 runs with different random seeds.

The primary comparison focused on predictive power. The four principal components from PCA yielded a model with an $R^2$ of 0.19. The latent space from the regular autoencoder performed similarly, resulting in an average $R^2$ of 0.15 (SD 0.01). In sharp contrast, our proposed method achieved a significantly higher average $R^2$ of 0.40 (SD 0.03), underscoring the benefit of guiding the latent space construction to capture outcome-related information.

Beyond its superior predictive power, our composite loss function also leads to a more stable latent representation regarding diverging patients. To assess this, we extended the stability analysis used for the subgroups to the entire latent dimension level. For each of the 15 runs, all patients were ranked based on their deviation from the global model for a specific latent dimension. The standard deviation of each patient's rank was then calculated across all runs, with a lower value indicating a more stable and consistent ordering of patients by that dimension. Our approach consistently produced more stable dimensions than the regular autoencoder. For instance, the latent dimension associated with dynamic vital capacity achieved a mean rank standard deviation of 26.1, a significant improvement over the 44.6 from the regular autoencoder. This enhanced stability was observed across all latent dimensions when aligned to the same underlying clinical construct across seeds.

These results highlight the advantage of our approach. While standard methods can effectively reduce dimensionality, their resulting latent spaces are not as informative for outcome prediction. By explicitly incorporating the prediction task into the loss function, our method creates a low-dimensional representation that is not only more predictive but also more stable in contrasting global and local models.

\section{Discussion} \label{sec:discussion}

In this study, we introduced a novel approach for contrasting global and localized regression modeling, specifically within a unified, low-dimensional latent space learned by an autoencoder. The primary application of our framework is to serve as a diagnostic tool. It evaluates when a global model might be inadequate by contrasting it with patient-specific alternatives. The global model constructed in our latent space acts as a relevant proxy for the simple, interpretable regression models that are a popular tool in biostatistics. By diagnosing the shortcomings of this common global model, we intend for our tool to be practically relevant in a familiar modeling context. This directly addresses an important challenge in biostatistics on how to move beyond a 'one-size-fits-all' model when subgroups are not known a priori and the data is complex. 

Central to our approach is the integration of an autoencoder for dimensionality reduction with patient-specific localized regression models. This integration was achieved through an end-to-end training process that minimizes a composite loss function that incentivizes both data reconstruction and revelation of local outcome-related associations. The impact of this outcome-aware dimensionality reduction strategy was evident when compared to standard methods. Our approach yielded a latent space with superior predictive power for the clinical outcome ($R^2=0.40$) than those derived from PCA ($R^2=0.19$) or a regular autoencoder trained solely for reconstruction ($R^2=0.15$). This quantitatively demonstrates that a latent space purposefully guided by the downstream prediction task, as achieved by our composite loss, is substantially more effective at uncovering clinically relevant patterns than spaces formed by techniques focused merely on data compression or maximal variance capture.

Our methodology can be viewed as an extension of exploratory techniques used in subgroup analysis and personalized medicine. While traditional subgroup analyses often rely on pre-defined covariates or recursive partitioning methods, our approach allows for the discovery of subgroups based on more complex patterns in a reduced-dimensional space that is optimized for this discovery. As a result, our method can help to group related variables into clinically meaningful latent dimensions, addressing the challenge of interpreting patterns in complex clinical datasets. For example, in the application to COPD data, our method effectively mitigated the curse of dimensionality by reducing the space spanned by patient characteristics from 76 dimensions to a four-dimensional latent space, thus enabling robust localized regression even in a setting where localization would not be feasible otherwise. As a result, our approach allows for a seamless comparison between global and localized models, facilitating the identification of individual patients and regions in the space of patient characteristics with significant deviations from the global model.

Applying our framework to the COPD clinical trial data demonstrated its effectiveness in balancing global and localized modeling. The global latent regression model adequately captured the outcome for the majority of patients. However, our method successfully identified two distinct patient groups with localized regression coefficients that significantly diverged from those of the global model. These subsets were characterized by distinct latent dimensions, confirming that the latent space preserved meaningful outcome-related structures. For both groups, the localized models provided a substantially greater reduction in prediction error compared to the modest improvement seen for the rest of the cohort, confirming that these were indeed patients for whom a personalized model was most advantageous.

By projecting these deviations back to the original predictor space, we translated latent patterns into interpretable findings. The identified groups exhibited distinct characteristics and showed significant interaction effects in the original predictors, underlining that the global model did not accurately capture their clinical profiles. This ability to connect patterns from the latent space to the original data suggests that our end-to-end learning approach effectively learned a low-dimensional representation that is meaningful and preserves the local relation to the outcome. 

Moreover, the consistency of results in the test data, despite its smaller sample size, supports the stability of the latent representation and the localized modeling within this low-dimensional space. This consistency enhances confidence in the reproducibility of our findings and suggests that our approach can generalize effectively to new samples from the same underlying population. 

When contrasted with a standard stepwise regression, both approaches identified the prominent interaction effects that defined our first subgroup, confirming a shared sensitivity for strong, global signals. Crucially, however, our approach also uncovered a relevant subgroup defined by localized gas-trapping effects (Group 2) that the stepwise procedure missed entirely. This highlights our approach's main advantage with its ability to serve as a diagnostic tool that detects important heterogeneity that is overlooked by conventional methods requiring strong global effects to select a variable.

Future research could explore the applicability of the proposed approach to settings beyond clinical prediction models, e.g. settings where the focus is on individualized treatment effects. This would require a modified loss function and a stronger focus on valid statistical inference, where the focus in the present setting had been on exploratory analyses. Relatedly, while our current method effectively identifies patient deviations using a heuristic based on global model confidence intervals, future work could advance from this exploratory flagging to formal statistical inference. This would require developing methods to robustly quantify the uncertainty of the patient-specific local model coefficients and, more importantly, the uncertainty of their deviation from the global effect. Such an extension would allow for hypothesis testing on whether a patient's local model is significantly different from the global one, strengthening the statistical foundation of the identified subgroups and their clinical interpretation. Furthermore, while our diagnostic tool is demonstrated within a regression framework, we hypothesize that the subgroups it identifies may also be the same individuals who would benefit from more complex, personalized approaches using other types of models. Additionally, a natural extension of our method would be to allow different outcome types, such as categorical and time-to-event data. We anticipate that a corresponding modification of the loss function would be straightforward.

Generally, the proposed approach is not intended to replace but to complement data analysis with global models. By seamlessly integrating dimensionality reduction with patient-specific regression, our method offers a diagnostic tool for assessing the adequacy of global models and identifying patients who may benefit from personalized modeling approaches. Such a combination of local with global approaches enhances model interpretability and flexibility, aligning with the growing emphasis on individualized analysis in clinical research. 

\vspace*{1pc}

\noindent {\bf{Acknowledgement}} This work was funded by the Deutsche Forschungsgemeinschaft (DFG, German Research Foundation) – Project-ID 499552394 – SFB 1597 (GB, AR, MH, HB, and MB). JMN is funded by the Berta-Ottenstein-Programme for Clinician Scientists, Faculty of Medicine, University of Freiburg.

\vspace*{1pc}

\noindent {\bf{Conflict of Interest}}

\noindent {\it{The authors have declared no conflict of interest.}}

\vspace*{1pc}

\noindent {\bf{Data availability statement}}

\noindent {The data are not publicly available due to privacy restrictions.}

\bibliographystyle{elsarticle-harv}
\bibliography{main}

\section*{Appendix}

\subsection*{A.1.\enspace Model Hyperparameters}

In addition to designing the autoencoder architecture, we selected several hyperparameters required for training. Hyperparameters were tuned through a grid search to optimize $\text{Loss}_{\text{pred}}$ while minimizing any significant degradation in reconstruction quality, measured by $\text{Loss}_{\text{rec}}$. To quantify the potential loss in reconstruction quality, we set $\lambda_{\text{pred}}$ to 0 and compared the resulting reconstruction losses.
For the autoencoder, the best performance was achieved with $\lambda_{\text{rec}}$ set to 1 and $\lambda_{\text{pred}}$ set to 0.06. The model was trained for 300 epochs using one batch.
For the localized regression, key hyperparameters included $\sigma$, which controls the width of the Gaussian kernel, and $k$-nearest, which adapts the kernel to the $k$-th nearest neighbor. We set $\sigma$ to 1.0 and defined $k$-nearest relative to the sample size, assigning it to 10\% of the samples, which resulted in approximately $k = 18$.

The computational demands for training the model with these hyperparameters and our application's dataset size ($n=173$ training samples, $p=76$ initial predictors, 300 epochs) are modest. The entire training process typically completed within 5 to 15 minutes on a standard contemporary desktop computer or laptop equipped with a multi-core CPU. This demonstrates that the method, for analyses of similar scale, is practically implementable without requiring specialized high-performance computing resources (e.g., dedicated GPUs), though leveraging such resources could offer further acceleration, particularly for larger datasets or more extensive hyperparameter searches.

\subsection*{A.2.\enspace Data Pre-Processing}

To ensure effective training of the autoencoder, we pre-processed the original dataset, which initially contained over 400 variables. First, we retained only variables with a variance greater than 0.2, removing those that contributed little information. This step reduced the dataset to 76 predictors, which served as input for the autoencoder. Next, we removed outliers, defined as values beyond four times the interquartile range, resulting in a final cohort of 217 patients for this study. Finally, before training, we standardized both the predictors and the outcome to have a mean of 0 and a standard deviation of 1.

\subsection*{A.3.\enspace Clustering Variables and Naming Clusters}

The Z-score profiles shown in Panel C of Figures \ref{fig:group1} and \ref{fig:group2} are based on clusters derived from hierarchical clustering of the predictors. Clustering was performed using absolute correlations between predictors and average linkage as the linkage criterion. The complete mapping of the 76 variables to the 21 clusters is available in the GitHub repository. Cluster names were assigned through a manual review of variable descriptions, ensuring that each name reflects the primary themes and clinical significance of the associated variables.

\subsection*{A.4.\enspace Naming Latent Dimensions}

We labeled each latent dimension according to the variables most strongly associated with it. To identify these associations, we split each latent dimension into two groups, above or below its median, then performed two-sample t-tests on every observed variable. We selected the ten variables with the highest absolute t-statistic for each dimension (Tables \ref{tab:latent_1_2} \& \ref{tab:latent_3_4}). These results are from the representative seed out of 15 seeds. 

Latent 1 is tied to expiratory airflow obstruction and diffusion capacity, reflecting the efficiency of gas exchange and airflow out of the lungs. Latent 2 highlights dynamic vital capacity, capturing the lung’s capacity to efficiently expel air. Latent 3 represents key expiratory lung volumes, capturing the total amount of air exhaled in measurements like FVC and FEV. Latent 4 focuses on static lung volumes, such as total lung capacity (TLC) and intrathoracic gas volume (ITGV), which are clinical indicators of gas trapping and hyperinflation

\begin{table}[h!]
\centering
\caption{Top ten variables most strongly associated with Latent 1 and Latent 2, ranked by absolute t-statistics.}
\label{tab:latent_1_2}
\begin{tabular}{ll}
\toprule
\textbf{Latent 1} & \textbf{Latent 2} \\
\midrule
FEV/VC \% Pre-B (t=-11.45) & VC \% Post-B (t=-12.57) \\
FEV/VC \% Post-B (t=-10.77) & VC \% Pre-B (t=-11.54) \\
ERV \% Pre-B (t=9.34) & FVC \% Post-B (t=-10.93) \\
MEF 75 \% Post-B (t=-9.24) & FVC \% Pre-B (t=-10.69) \\
MEF 75 \% Pre-B (t=-9.14) & VCMAX \% Pre-B (t=-10.60) \\
ERV Liters Pre-B (t=8.79) & VCMAX \% Post-B (t=-10.24) \\
MEF 50 \% Post-B (t=-8.08) & ERV \% Pre-B (t=-8.02) \\
MEF 75 L/s Post-B (t=-8.03) & ERV \% Post-B (t=-6.94) \\
MEF 75 L/s Pre-B (t=-8.00) & BORG Scale (t=5.10) \\
DLCO/VA mmol/min/kPa/L Post-B (t=-7.98) & Weight kg (t=4.90) \\
\bottomrule
\end{tabular}
\end{table}

\begin{table}[h!]
\centering
\caption{Top ten variables most strongly associated with Latent 3 and Latent 4, ranked by absolute t-statistics.}
\label{tab:latent_3_4}
\begin{tabular}{ll}
\toprule
\textbf{Latent 3} & \textbf{Latent 4} \\
\midrule
VC Liters Pre-B (t=-13.47) & ITGV \% Post-B (t=-13.16) \\
VC Liters Post-B (t=-13.30) & TLC \% Post-B (t=-12.66) \\
VCMAX Liters Pre-B (t=-13.16) & ITGV \% Pre-B (t=-12.45) \\
VA Liters Post-B (t=-12.95) & TLC \% Pre-B (t=-10.95) \\
FVC Liters Post-B (t=-12.82) & RV \% Post-B (t=-10.38) \\
VCMAX Liters Post-B (t=-12.77) & RV \% Pre-B (t=-10.28) \\
FVC Liters Pre-B (t=-12.54) & RV/TLC \% Pre-B (t=-10.16) \\
FEV Liters Post-B (t=-9.21) & FEV Liters Pre-B (t=9.95) \\
PEF L/s Pre-B (t=-9.04) & FEV Liters Post-B (t=9.91) \\
FEV Liters Pre-B (t=-9.04) & PEF L/s Post-B (t=9.90) \\
\bottomrule
\end{tabular}
\end{table}

Table \ref{tab:correlation_matrix} shows that all dimensions are relatively uncorrelated. To promote more distinct patterns, we regularized the correlation between latent dimensions by adding a penalty term on the squared Pearson correlation between each pair of latent dimensions:

$$
\text{Loss}_{\text{reg}} = \sum_{k \neq l}^d \text{corr}(z_k, z_l)^2,
$$

where $\text{corr}(z_k, z_l)$ is the Pearson correlation between the $k$-th and $l$-th latent dimensions, and $d$ is the total number of dimensions. We assigned this regularization a weight of $\lambda_{\text{reg}} = 0.3$. This penalty helped keep the dimensions relatively uncorrelated, improving interpretability.

\begin{table}[ht]
    \centering
    \caption{Correlation Matrix of Latent Dimensions}
    \begin{tabular}{lcccc}
        \toprule
        & Latent 1 & Latent 2 & Latent 3 & Latent 4 \\
        \midrule
        Latent 1 & 1.00  & -0.03 & -0.01 & -0.04 \\
        Latent 2 & -0.03 & 1.00  &  0.01 &  0.04 \\
        Latent 3 & -0.01 & 0.01  &  1.00 & -0.02 \\
        Latent 4 & -0.04 & 0.04  & -0.02 &  1.00 \\
        \bottomrule
    \end{tabular}
    \label{tab:correlation_matrix}
\end{table}

\subsection*{A.5.\enspace Full Stepwise Regression Models}

As mentioned in Section 3.5, a standard stepwise regression procedure was performed on the original 76 predictors to contrast with our proposed method. The procedure was run using different p-value thresholds for the initial univariate screening step. The final models selected by the backward and forward AIC-based procedure for each threshold are detailed below.

\begin{table}[ht]
\centering
\caption{Final Stepwise Regression Model for Univariate Screening with $p < 0.05$. $R^2 = 0.57$.}
\label{tab:stepwise_05}
\begin{tabular}{lrrrrrr}
\toprule
\textbf{Variable} & \textbf{Coef.} & \textbf{Std. Err.} & \textbf{t} & \textbf{$\mathbf{P>|t|}$} & \textbf{[0.025} & \textbf{0.975]} \\
\midrule
Intercept & -0.08 & 0.06 & -1.26 & 0.21 & -0.20 & 0.04 \\
COPD Symptoms & 0.08 & 0.06 & 1.39 & 0.17 & -0.04 & 0.20 \\
Breathlessness Score & 0.01 & 0.06 & 0.14 & 0.89 & -0.11 & 0.13 \\
FVC Liters (Pre-B) & 0.19 & 0.08 & 2.42 & 0.02 & 0.03 & 0.34 \\
FEV \% Pre-B & -0.48 & 0.18 & -2.73 & 0.01 & -0.83 & -0.13 \\
VCMAX \% Pre-B & -0.18 & 0.08 & -2.22 & 0.03 & -0.33 & -0.02 \\
FVC \% Pre-B & -0.17 & 0.09 & -1.82 & 0.07 & -0.36 & 0.02 \\
FEV \% Post-B & 0.55 & 0.19 & 2.94 & 0.00 & 0.18 & 0.92 \\
DLCO \% Pre-B & -0.15 & 0.08 & -1.96 & 0.05 & -0.31 & 0.00 \\
6MWT Distance & -0.11 & 0.07 & -1.66 & 0.10 & -0.24 & 0.02 \\
BORG Scale & 0.51 & 0.07 & 7.67 & 0.00 & 0.38 & 0.64 \\
Breathlessness Score $\times$ VCMAX \% Pre-B & -0.16 & 0.05 & -3.19 & 0.00 & -0.26 & -0.06 \\
Breathlessness Score $\times$ FEV \% Post-B & 0.19 & 0.06 & 3.27 & 0.00 & 0.08 & 0.30 \\
Breathlessness Score $\times$ FVC Liters Pre-B & -0.19 & 0.06 & -2.95 & 0.00 & -0.31 & -0.06 \\
VCMAX \% Pre-B $\times$ FVC \% Pre-B & 0.09 & 0.04 & 2.18 & 0.03 & 0.01 & 0.18 \\
\bottomrule
\end{tabular}
\end{table}

\begin{table}[ht]
\centering
\caption{Final Stepwise Regression Model for Univariate Screening with $p < 0.10$. $R^2 = 0.59$.}
\label{tab:stepwise_10}
\begin{tabular}{lrrrrrr}
\toprule
\textbf{Variable} & \textbf{Coef.} & \textbf{Std. Err.} & \textbf{t} & \textbf{P$>|$t$|$} & \textbf{[0.025} & \textbf{0.975]} \\
\midrule
Intercept & 0.01 & 0.06 & 0.12 & 0.91 & -0.11 & 0.13 \\
COPD Symptoms & 0.16 & 0.06 & 2.68 & 0.01 & 0.04 & 0.28 \\
VCMAX \% Pre-B & -0.15 & 0.11 & -1.44 & 0.15 & -0.37 & 0.06 \\
FVC Liters & 0.15 & 0.11 & 1.26 & 0.21 & -0.08 & 0.37 \\
FEV \% Pre-B & -0.49 & 0.20 & -2.46 & 0.01 & -0.88 & -0.09 \\
PEF \% Pre-B & -0.09 & 0.08 & -1.09 & 0.28 & -0.26 & 0.08 \\
VC \% Pre-B & 0.14 & 0.13 & 1.09 & 0.28 & -0.12 & 0.40 \\
VCMAX \% Post-B & -0.21 & 0.08 & -2.55 & 0.01 & -0.38 & -0.05 \\
FVC \% Post-B & -0.18 & 0.13 & -1.39 & 0.17 & -0.43 & 0.08 \\
FEV \% Post-B & 0.76 & 0.21 & 3.71 & 0.00 & 0.36 & 1.17 \\
RV \% Post-B & 0.41 & 0.13 & 3.15 & 0.00 & 0.15 & 0.67 \\
VCL \% Pre-B & -0.08 & 0.13 & -0.59 & 0.56 & -0.34 & 0.18 \\
RV/TLC \% Post-B & -0.39 & 0.16 & -2.44 & 0.02 & -0.71 & -0.07 \\
DLCO \% Pre-B & -0.18 & 0.08 & -2.29 & 0.02 & -0.34 & -0.03 \\
6MWT Distance & -0.17 & 0.07 & -2.52 & 0.01 & -0.30 & -0.04 \\
BORG Scale & 0.53 & 0.07 & 8.09 & 0.00 & 0.40 & 0.66 \\
COPD Symptoms $\times$ VCMAX \% Post-B & 0.17 & 0.06 & 2.80 & 0.01 & 0.05 & 0.29 \\
COPD Symptoms $\times$ VCL \% Pre-B & -0.16 & 0.06 & -2.51 & 0.01 & -0.28 & -0.03 \\
FVC Liters $\times$ FEV \% Post-B & -0.26 & 0.07 & -3.61 & 0.00 & -0.40 & -0.12 \\
PEF \% Pre-B $\times$ FVC \% Post-B & 0.20 & 0.07 & 2.71 & 0.01 & 0.05 & 0.35 \\
FVC Liters $\times$ 6MWT Distance & 0.11 & 0.05 & 2.18 & 0.03 & 0.01 & 0.21 \\
\bottomrule
\end{tabular}
\end{table}

\begin{table}[ht]
\centering
\caption{Final Stepwise Regression Model for Univariate Screening with $p < 0.157$. $R^2 = 0.52$.}
\label{tab:stepwise_157}
\begin{tabular}{lrrrrrr}
\toprule
\textbf{Variable} & \textbf{Coef.} & \textbf{Std. Err.} & \textbf{t} & \textbf{$\mathbf{P>|t|}$} & \textbf{[0.025} & \textbf{0.975]} \\
\midrule
Intercept & 0.03 & 0.06 & 0.59 & 0.56 & -0.08 & 0.14 \\
COPD Symptoms & 0.11 & 0.06 & 1.88 & 0.06 & -0.01 & 0.23 \\
VCMAX \% Pre-B & -0.19 & 0.07 & -2.77 & 0.01 & -0.33 & -0.05 \\
FEV/VC \% Pre-B & -0.62 & 0.18 & -3.42 & 0.00 & -0.98 & -0.26 \\
PEF \% Pre-B & -0.13 & 0.08 & -1.74 & 0.08 & -0.29 & 0.02 \\
RV \% Pre-B & -0.13 & 0.09 & -1.51 & 0.13 & -0.31 & 0.04 \\
FEV/VC \% Post-B & 0.74 & 0.18 & 4.04 & 0.00 & 0.38 & 1.10 \\
RV \% Post-B & 0.37 & 0.13 & 2.91 & 0.00 & 0.12 & 0.62 \\
RV/TLC \% Post-B & -0.31 & 0.12 & -2.65 & 0.01 & -0.54 & -0.08 \\
6MWT Distance & -0.18 & 0.07 & -2.74 & 0.01 & -0.32 & -0.05 \\
BORG Scale & 0.57 & 0.07 & 8.61 & 0.00 & 0.44 & 0.69 \\
RV \% Post-B $\times$ 6MWT Distance & 0.11 & 0.05 & 1.98 & 0.05 & 0.00 & 0.22 \\
\bottomrule
\end{tabular}
\end{table}

\end{document}